# Status of the MARS code

*Igor* Rakhno [1*], *Igor* Tropint[1], *Dali* Georgobiani[1] and *Alajos* Makovec[1]

[1]Fermi National Accelerator Laboratory, P.O. Box 500, Batavia, Illinois 60510-5011, USA

**Abstract.** This report describes major features of the most recent version of the MARS code as well as ongoing developments. The list of features includes various options for geometry models, a beam line builder based on MADX code, import of geometry models in GDML format, use of structured and unstructured meshes for scoring purposes, an update to the recent TENDL library for a number of projectiles at low energies (up to 250 MeV), and a recently implemented method to calculate spatial distribution of residual dose in a single computer run without an intermediate source. Examples of the code application to various projects are presented as well.

## 1 Introduction

MARS code is a general-purpose radiation transport Monte Carlo code with an emphasis on accelerator applications [1-7]. It provides detailed simulation of hadronic and electromagnetic cascades in an arbitrary three-dimensional (3D) geometry of shielding, accelerator, detector and spacecraft components in a broad energy range: 1 keV to 100 TeV for hadrons, muons, heavy ions and electromagnetic showers, and $10^{-5}$ eV to 100 TeV for neutrons. All interactions in the entire energy range can be simulated either inclusively or exclusively using a combination of event generators. Nuclide production, decay, transmutation and calculation of residual activity can be done with the built-in DeTra code [8] or using SandiaDecay library [9]. The ENDF/B-VIII.0 evaluated nuclear data library [10] is used to process neutron interactions with matter below 14 MeV. Given that the library provides the nuclear data for every single isotope—instead of the previous approach focused on materials as natural mixtures—detailed isotope composition for natural mixtures is employed to process low energy neutron interactions with matter. Radiation damage cross sections for neutrons are derived from the ENDF/B-VIII.0 library with NJOY2016 code [11]. Geometry models can be described in different ways which provides essential flexibility. MARS15 code is routinely used with ANSYS code to study thermal and mechanical properties of materials under irradiation. The following sections describe features related to recent updates.

## 2 Geometry models

Various types of geometry model description are available in MARS15 code. The major advantage is in using a so-called *Non-standard* geometry option which has the highest priority. This option allows us to implement parametric additions and/or corrections to an initial complex model developed with any other geometry. The *Non-standard* geometry can be implemented by users in different ways depending on their preferences and available files. For example, it can be implemented as a Fortran file, or a C++ file which employs ROOT [12] geometry, or a GDML file. Below we describe major existing geometry features.

### 2.1 Basic constructive solid geometry (CSG)

The basic CSG geometry in MARS15 code—a so-called *Extended geometry*—is a very convenient tool to develop a geometry model for relatively simple and moderately complex structures (up to a few hundred objects) including beam lines without bending magnets. For more complicated structures including beam lines with bending magnets—in both horizontal and vertical planes—a different approach based on ROOT geometry is preferable.

### 2.2 ROOT geometry

The ROOT geometry package features about twenty basic shapes (primitives) that allow us to build complex geometry models as well as perform particle tracking and visualization. Composite shapes can be built out of the primitives using various translation and rotation matrices as well as Boolean operations. The shapes can be divided into slices as well as replicated at other locations. In order to create such geometry models, users need to work with source code in C++. At present, there are no options to create such models using MARS input files or graphical user interface (GUI). However, there is a third-party software that allows us to perform such a model development in a GUI [13].

Examples of models developed using the ROOT geometry option are shown in Figures 1 and 2. Also,

---
* Corresponding author: rakhno@fnal.gov

these Figures show beam lines built with the beam line builder based on MADX code (see Sec. 3.1 below).

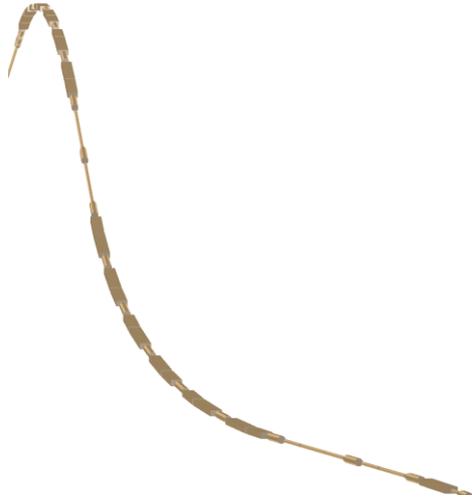

**Fig. 1.** A fragment of a 3D beam line model with bending magnets in both horizontal and vertical planes. The model has been built using ROOT geometry and the MARS15 beam line builder based on MADX code.

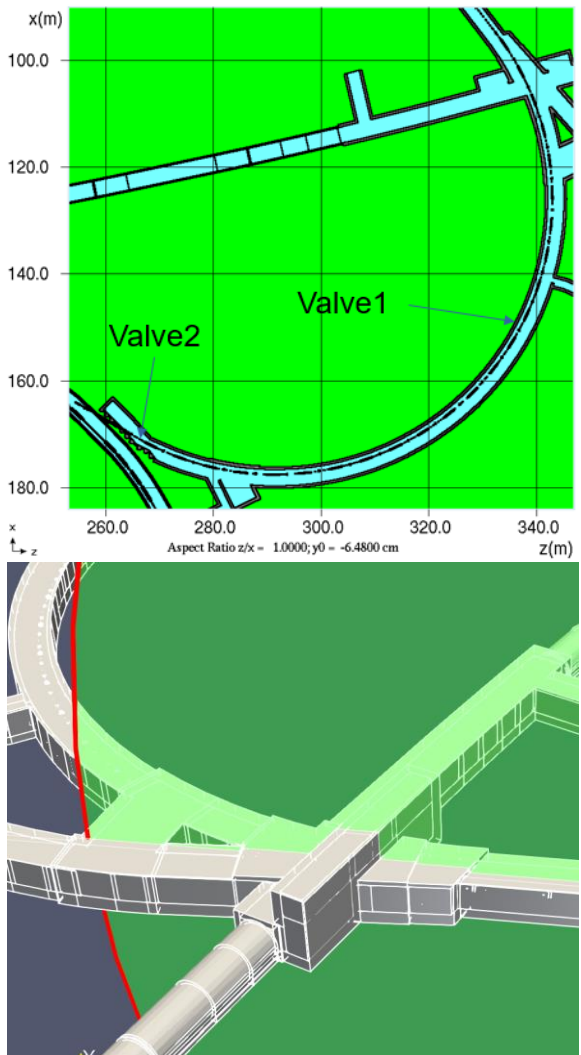

**Fig. 2.** A fragment of a beam line in a tunnel with other tunnels crossing the first one. The model has been build with ROOT geometry and MARS15 beam line builder based on MADX code. The fragment is shown in two dimensions (2D) and as a 3D view (top and bottom, respectively).

## 2.3 Import of geometry models in GDML format

The Geometry Description Markup Language (GDML) is an application-independent geometry description format based on Extensible Markup Language (XML). The latter enables us to store data in plain text format which facilitates storing and sharing the data. The GDML format is used by Geant4 code [14], so that import of GDML models into MARS15 code enables us to perform straightforward comparisons between the two codes. In MARS15 code, a GDML geometry model is treated as a model written in *Non-standard* geometry, and the import process is complemented by a proper material assignment to every single geometry region according to MARS15 rules. The latter is performed using an in-house routine.

## 2.4 Meshes and scoring

At present, both 2D and 3D meshes can be used in MARS15 code for scoring purposes. Various 2D distributions can be stored in both HBOOK and ROOT formats. Visualization of the calculated distributions of physical quantities is performed with MARS15 GUI. A verification between distributions in the older and new formats—HBOOK and ROOT, respectively—is currently underway. After the comparison process completes, only 2D distributions of physical quantities in the newer ROOT format will be kept.

The 3D meshes and calculated distributions of various physical quantities are presented in VTK format [15]. Visualization of the calculated distributions in such a case can be performed using a GUI provided by ParaView software [16].

It should be noted that in both the cases of 2D and 3D meshes, the geometry model itself is independent of the meshes. At the same time, contemporary Computer-Aided Design (CAD) software can provide very detailed models of realistic complex systems such as accelerator facilities with corresponding infrastructure. Therefore, from practical standpoint it is very important to employ in radiation transport calculations the (almost) very same model that can be produced with a CAD software. In this regard, there are various roadblocks, such as overwhelming number of small parts in the CAD model or overlaps between neighbour parts of the model. However, the possibility to use (almost) the same CAD model for radiation transport calculations as well as for thermal and mechanical studies (using MARS and ANSYS, respectively) can improve quality of results and save significant time due to elimination of need for model developments from scratch. At the least, such a possibility means that instead of time spent on model development for a radiation transport code (human time) one may has to deal with increased time spent on calculations (computer time).

In order to investigate feasibility and quality of such an approach using MARS15 code, energy deposition calculations have been performed for a target designed for the MiniBooNE experiment with an 8-GeV proton beam at Fermilab [17] (see Figure 3). The following procedures have been used for the calculations (see Figure 4):

- A ROOT geometry model has been developed.
- The geometry model from MARS code has been converted to a CAD format (STEP) using an in-house procedure.
- ANSYS code generated an unstructured mesh using the STEP file from MARS code, and the mesh has been exported in CGNS format.
- The spatial distribution of absorbed energy in the target has been calculated with MARS code on the unstructured mesh from ANSYS.
- For subsequent thermal and mechanical analysis with the ANSYS code, the same unstructured mesh can be used.

As a consequence of such an approach, boundaries of the mesh cells completely coincide with geometric boundaries of the cells.

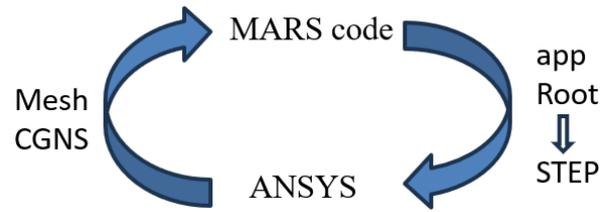

**Fig. 4.** A workflow for radiation and thermal studies with the same mesh using MARS and ANSYS, respectively.

A 3D cut away view of the MARS model of the MiniBooNE target assembly, including a CGNS mesh overlayed over the geometry model, is shown in Figure 5. Calculated distributions of energy deposition in the target assembly are shown in Figure 6. One can see that the distribution, calculated in the target along the beam axis using the unstructured 3D mesh, provides a high-quality result with minor irregularities—as large as a few percent or less—observed at transition regions between different longitudinal target sections. At the same time, regular approach without using CAD meshes reveals much larger irregularities at the transition regions where different materials can belong to the same voxel of the regular non-CAD mesh.

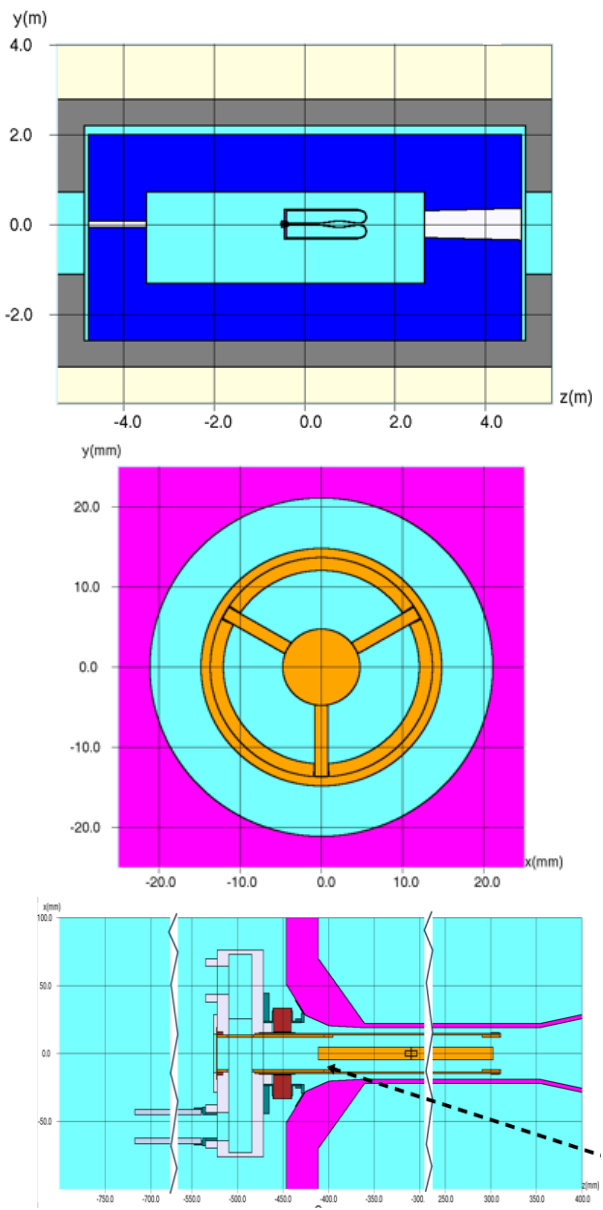

**Fig. 3.** Target hall for MiniBooNE experiment, pulsed horn and horn shielding (top); a cross section of the MiniBooNE target with upstream slug locator and locator fins (middle); a detailed plan view of MiniBooNE target assembly with the horn (bottom).

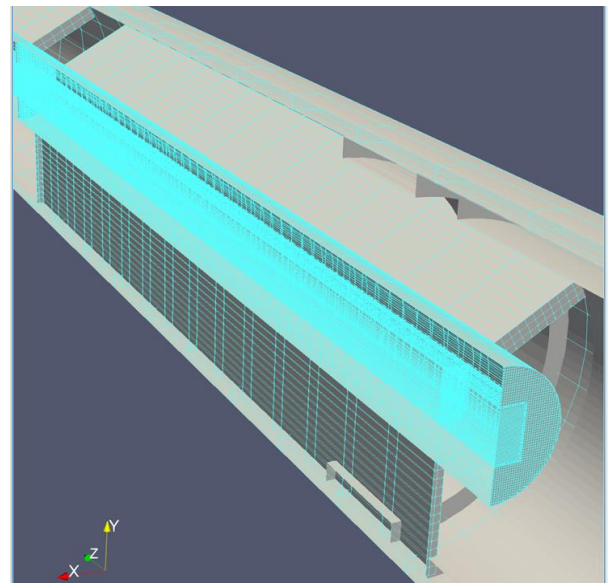

**Fig. 5.** A 3D cut away view of the MARS model of the MiniBooNE target assembly including a CGNS mesh overlayed over the geometry model.

## 3 Particle transport and interactions

In this Section, several recent updates, related to particle transport and interactions, are presented.

### 3.1 Beam line builder and tracker

In accelerator applications, an essential feature is a possibility to build models that follow realistic—often 3D—beamlines (see Figure 1). For MARS code, a beam line builder has been developed which employs information from MADX input. Usually, the latter becomes available after the corresponding beam line design stage is completed. When using the builder, all components of the geometry model—beam pipe, magnets, tunnel, penetrations and so on—follow the beam line (axis) which serves as a driver in the model. Examples of models developed with the builder are shown in Figures 2, 7 and 8.

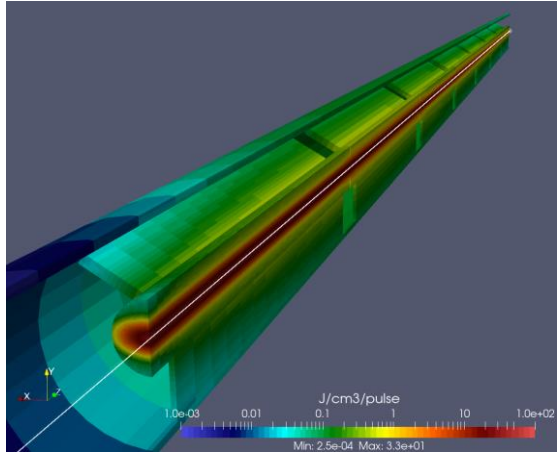

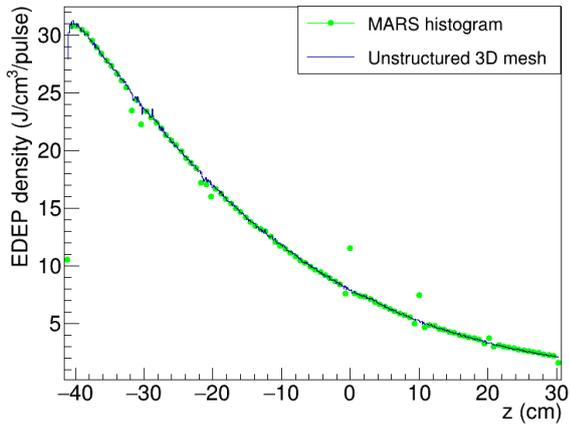

**Fig. 6.** A 3D distribution of energy deposition density in the MiniBooNE target assembly calculated using an unstructured grid obtained with ANSYS code (top) and the energy deposition density calculated in the target along the beam axis in comparison with results obtained using built-in MARS estimators (2D histograms) (bottom).

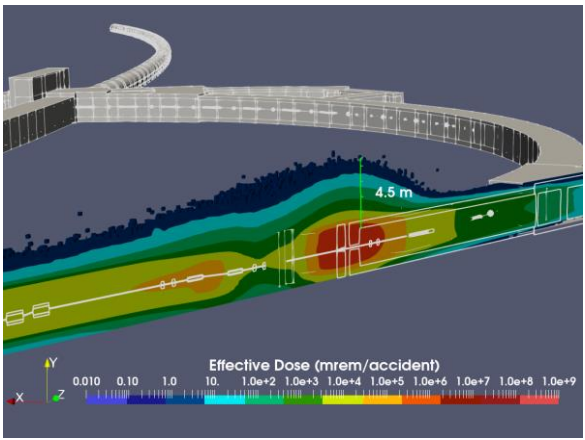

**Fig. 7.** A distribution of prompt dose rate, calculated for a beam accident on a magnet, overlayed over the corresponding 3D tunnel model implemented using the described above beam line builder. The Figure was prepared with ParaView [16].

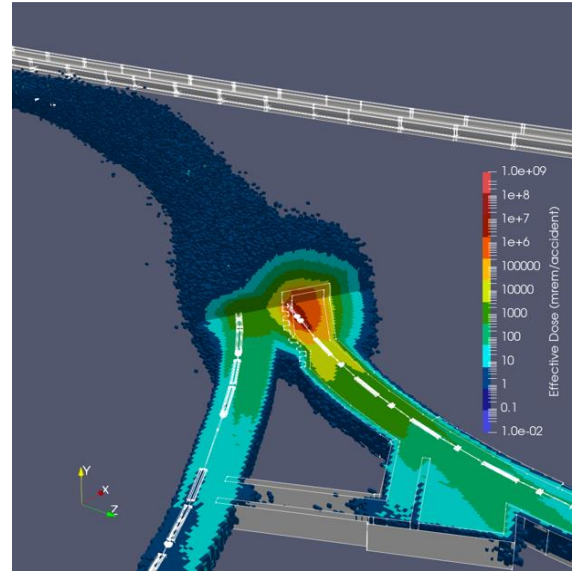

**Fig. 8.** A cut away view of prompt dose rate, calculated for a beam accident on a magnet, overlayed over the corresponding 3D tunnel model implemented using the described above beam line builder. The Figure was prepared with ParaView [16].

In order to further improve modelling of beam transport and interactions with matter for accelerator applications, a tracker (stepper) based on PTC tracker has been developed for MARS code [4]. It allows us to perform precise analytical calculations of beam transport in vacuum inside beam pipe aperture using the same distributions of electromagnetic field that were used for the beam line design. In matter outside the aperture, Monte Carlo modelling of beam transport and interactions is performed with MARS code. Exchange of particle phase space coordinates between PTC and MARS codes is performed at the border between vacuum and matter. Quality of the corresponding beam transport is shown in Figure 9 below.

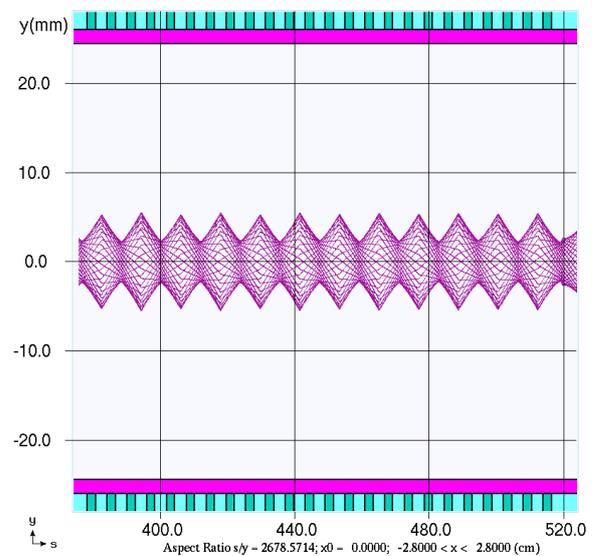

**Fig. 9.** An example of precise particle tracking in vacuum inside a beam pipe performed using the PTC tracker implemented into MARS code.

## 3.2 Interactions at low energies

The library of evaluated nuclear data TENDL [18] provides a comprehensive database for modelling interactions of several projectiles—p, n, d, t, $^3$He, $^4$He, and γ—with matter at energies from 1 MeV up to 200 MeV. In the most recent version, TENDL-2023, the lower and upper energies have been extended down to 250 keV and up to 600 MeV, respectively. Energy and angular distributions of all secondaries—p, n, d, t, $^3$He, $^4$He, γ as well as residual nuclei—are provided for all stable nuclei from periodic table of elements as well as for unstable nuclei as targets. An initial (inclusive) version of our in-house processing software and corresponding event generator were developed in 2014 and presented at SATIF-12. This activity was driven, most of all, by ongoing work on Proton Improvement Plan [19] at Fermilab. The update from TENDL-2021 to TENDL-2023 includes not only replacement of files but an extension to an exclusive option as well. At present, the inclusive option (s0) is being tested. Some comparisons between TENDL 2023 data—as modelled by MARS code—and experimental data for several projectile-target combinations are presented in Figures 10 and 11.

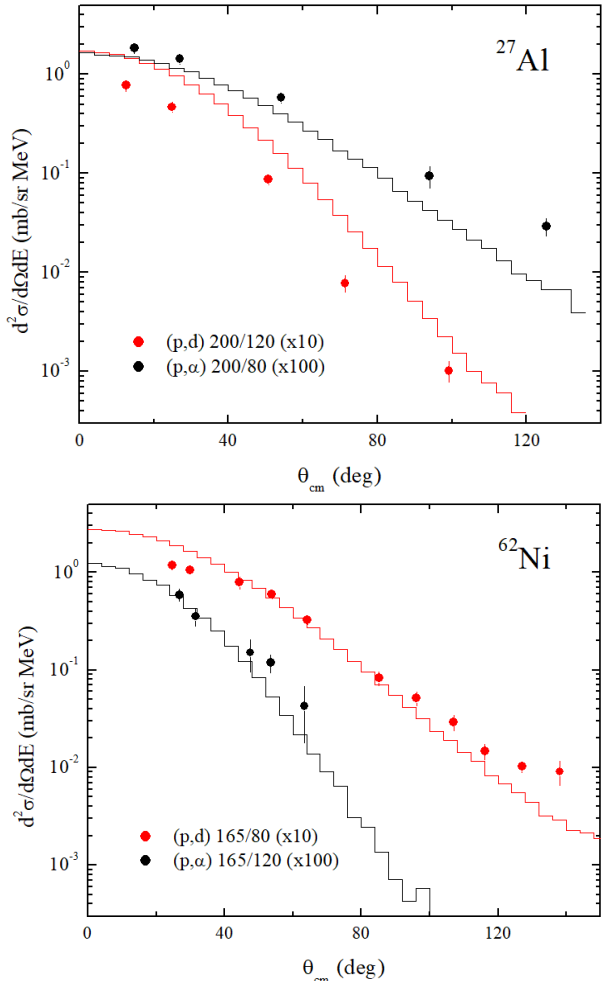

**Fig. 10.** Calculated with TENDL-2023 (histograms) and measured (symbols) energy and angular distributions of secondary deuterons and α-particles emitted from aluminium and nickel targets. The experimental data are from EXFOR database [20]. Incident proton energy was equal to 200 and 165 MeV for the $^{27}$Al and $^{62}$Ni targets, respectively. Energy of the emitted particles is indicated on the Figures in MeV.

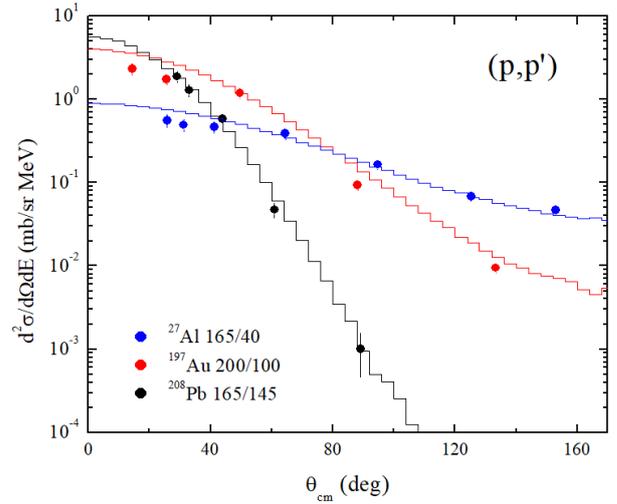

**Fig. 11.** Calculated with TENDL-2023 (histograms) and measured (symbols) energy and angle distributions of protons which experience inelastic scattering on three different target nuclei. The experimental data are from EXFOR database. Kinetic energy of the incoming and outgoing protons is shown on the Figure in MeV.

## 3.3 Spatial distribution of residual dose

Recently, a method has been implemented that allow us to calculate 3D distributions of residual dose around irradiated objects. It is based on a separate modelling performed for residual radiation emitted from unstable residual nuclei. This method complements calculation of 2D contact residual dose distributions used in MARS code previously. Various comparisons with experimental data from SINBAD database and with FLUKA calculations have been performed [21] (see Figures 12 and 13 as an example). One can see that, except for the case of the copper sample and cooling time less than 2 days, the agreement between the experimental data and calculations with FLUKA and MARS codes is excellent.

Calculated distributions of residual dose rate for LBNF Hadron Absorber [22] and PIP-II Instrumentation Cart [19] are shown in Figures 14 and 15, respectively. Such distributions are very helpful when planning various hands-on procedures during scheduled maintenance.

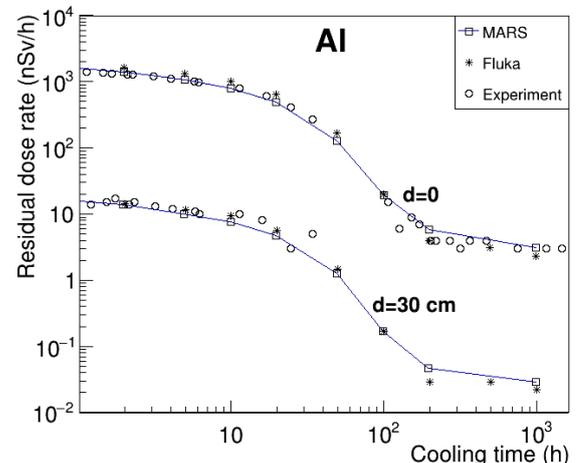

**Fig. 12.** Calculated and measured residual dose rate on contact and at 30 cm from the irradiated aluminium sample [21].

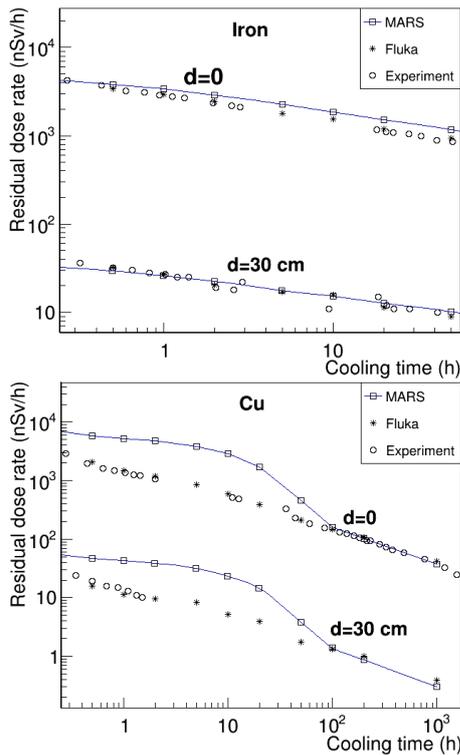

**Fig. 13.** The same as in Figure 12 but for iron (top) and copper (bottom) samples.

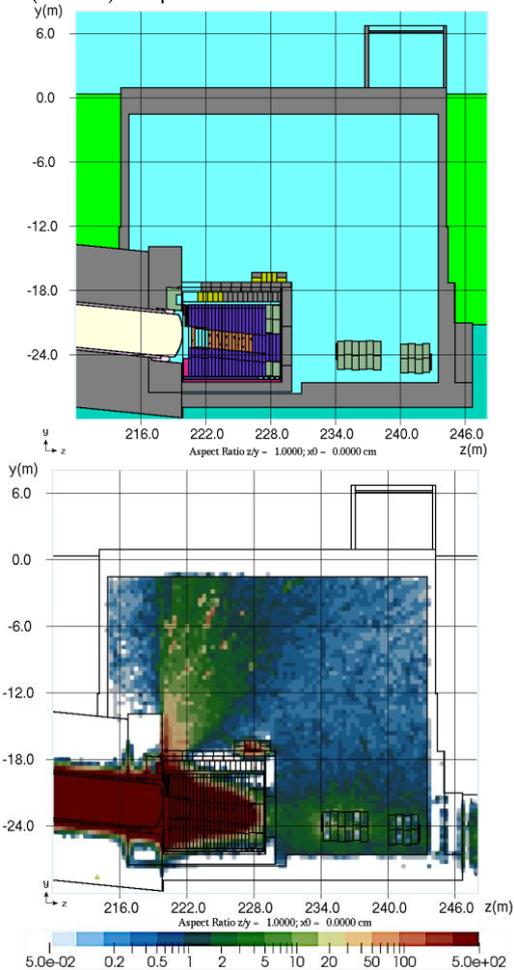

**Fig. 14.** An elevation view of MARS model of LBNF Hadron Absorber Complex [22] (top) and calculated distribution of residual dose rate around the Absorber for a maintenance scenario with seven concrete blocks removed (bottom). A continuous irradiation with a 120-GeV proton beam at 2.4 MW followed by 1-day cooling is assumed.

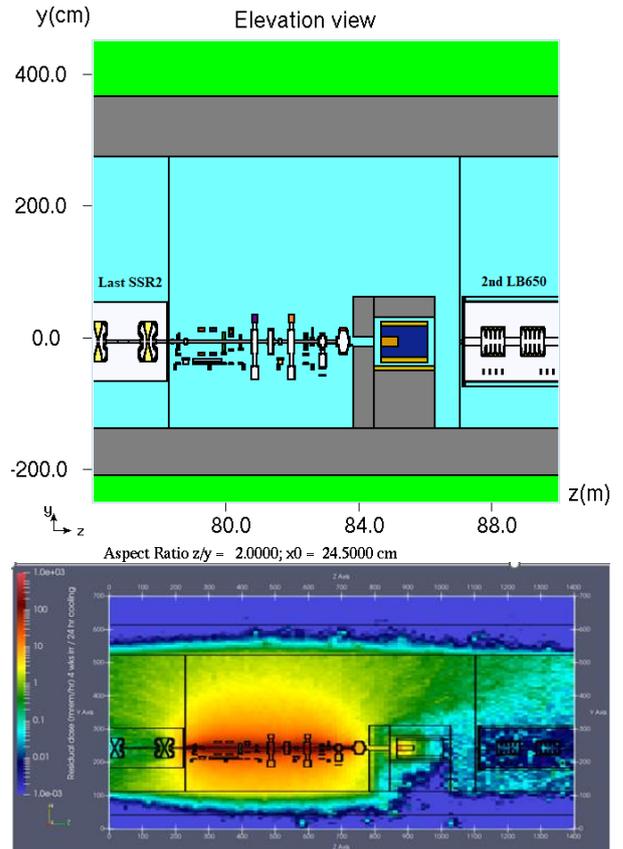

**Fig. 15.** A fragment of PIP-II Linac model with an Instrumentation Cart at 177 MeV location (top) and calculated distribution of residual dose rate for a commissioning and beam diagnostics scenario with beam loss rate on the 5-meter-long beam pipe fragment equal to 0.73 W. Irradiation during 4 weeks with subsequent cooling for 24 hours is assumed.

## 4 Work in progress

Work on development of a new more user-friendly GUI is underway. The new main window is planned to facilitate switching between different windows. For example, a material window (see Figure 16 below) will allow users to browse built-in elements and compounds, copy these into a material input file, import and export material input files, edit data for the compounds and so on.

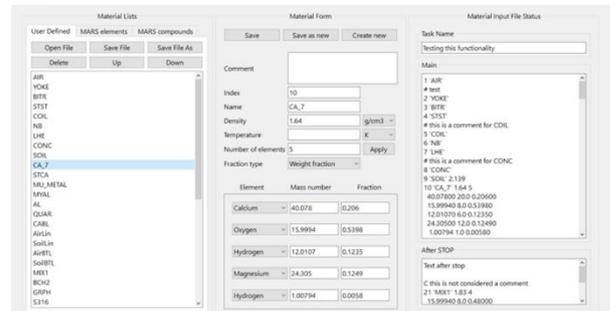

**Fig. 16.** A prototype of the new material window in GUI.

Also, work on replacement of obsolete features (for example, COMMON blocks in Fortran), removal of global static data and improved modularity is underway.

**High Energy Physics. Publisher acknowledges the U.S. Government license to provide public access under the DOE Public Access Plan.**

**This research used Fermigrid at the Fermi National Accelerator Laboratory and ALCC allocations at the Argonne Leadership Computing Facility (ALCF), which is a DOE Office of Science User Facility supported under Contract DE-AC02-06CH11357, and at the National Energy Research Scientific Computing Center (NERSC), which is a DOE Office of Science User Facility supported under Contract No. DE-AC02-05CH11231.**